\documentclass{article}

\PassOptionsToPackage{square, numbers}{natbib}
\usepackage[final]{neurips_2021_ml4ps}

\usepackage[utf8]{inputenc} % allow utf-8 input
\usepackage[T1]{fontenc}    % use 8-bit T1 fonts
\usepackage{hyperref}       % hyperlinks
\usepackage{url}            % simple URL typesettinghttps://www.overleaf.com/project/61294376a132855f47cf37c1
\usepackage{booktabs}       % professional-quality tableshttps://www.overleaf.com/project/61294376a132855f47cf37c1
\usepackage{amsfonts}       % blackboard math symbols
\usepackage{nicefrac}       % compact symbols for 1/2, etc.
\usepackage{microtype}      % microtypography
\usepackage{xcolor}         % colors
\usepackage{chemformula}
\usepackage{caption}
\usepackage{subcaption}

\title{Inorganic Synthesis Reaction Condition Prediction with Generative Machine Learning}

\author{
  Christopher Karpovich, Zach Jensen, Vineeth Venugopal, Elsa Olivetti \\
  Dept. of Materials Science and Engineering\\
  Massachusetts Institute of Technology\\
  Cambridge, MA 02139 \\
  \texttt{\{ckarp, zjensen, vineethv, elsao\}@mit.edu} \\
}

\begin{document}

\maketitle

\begin{abstract}
Data-driven synthesis planning with machine learning is a key step in the design and discovery of novel inorganic compounds with desirable properties. Inorganic materials synthesis is often guided by chemists' prior knowledge and experience, built upon experimental trial-and-error that is both time and resource consuming. Recent developments in natural language processing (NLP) have enabled large-scale text mining of scientific literature, providing open source databases of synthesis information of synthesized compounds, material precursors, and reaction conditions (temperatures, times). In this work, we employ a conditional variational autoencoder (CVAE) to predict suitable inorganic reaction conditions for the crucial inorganic synthesis steps of calcination and sintering. We find that the CVAE model is capable of learning subtle differences in target material composition, precursor compound identities, and choice of synthesis route (solid-state, sol-gel) that are present in the inorganic synthesis space. Moreover, the CVAE can generalize well to unseen chemical entities and shows promise for predicting reaction conditions for previously unsynthesized compounds of interest.
\end{abstract}

\section{Introduction}
Virtual materials screening and physics-based simulations have in recent years greatly accelerated the design and discovery of novel inorganic compounds with applications in chemical catalysis \citep{Toyao2019}, thermoelectrics \citep{Iwasaki2019}, and metal-organic frameworks \citep{fernandez2014rapid}. However, while existing tools have largely focused on the inverse design of inorganic materials, one major remaining challenge is the development of inverse synthesis planning, where, given a target material, appropriate synthesis parameters are suggested as a means to synthesize the compound. In comparison to widely available organic chemistry reaction databases, the vast majority of openly accessible inorganic synthesis information is contained within the text of scientific journal articles \citep{Olivetti2020}. Recent efforts have leveraged advances in Natural Language Processing (NLP) to extract and convert inorganic synthesis information in unstructured scientific text into machine readable databases \citep{Kimoxide, Kononova2019, Weston2019}. In the organic chemistry space, previous works have investigated synthesis temperature prediction using feedforward neural network based models \citep{Gao2018}; however, they utilize a dataset on the order of $10^{7}$ points with a small range of reported temperatures (-100 to 300 $^{\circ}$C), while inorganic synthesis datasets typically consist of $10^{4}$ to $10^{5}$ points and report temperatures ranging from 200 to 2000 $^{\circ}$C. The dearth of available inorganic synthesis data and wide range of reported reaction conditions makes the inorganic prediction problem challenging. In the inorganic space, other works have developed methods for condition generation for specific materials families such as \ch{TiO2}, \ch{MnO2}, and \ch{SrTiO3} \citep{Kim2017}, inverse prediction of precursor materials and synthesis operation sequences \citep{Kim2020}, forward prediction of target compositions \citep{Malik2021}, and precursor selection based on kinetic factors \citep{Aykol2021}; however, to the best of our knowledge, no studies have explored the prediction of synthesis conditions for novel inorganic compounds. 

Generative machine learning models have already proven to be powerful tools in the chemical and materials spaces for materials discovery. Autoencoder models have been leveraged to optimize molecules with desirable properties over a learned latent space \citep{Gomez-Bombarelli2018} and predict crystal structures for new inorganic compounds \citep{Court2020}. Generative adversarial networks have also been explored for drug discovery \citep{Guimaraes2017} and to screen inorganic material compositions \citep{Dan2020}. In the case of experimental synthesis, reaction conditions present several important variables in the validation of computer-aided synthesis planning. Broadly, in solid-state synthesis, two or more non-volatile solid precursor materials are ground and heated in multiple consecutive steps to temperatures below their melting points to react and form the desired product \citep{west2014solid, Chamorro2018}. In sol-gel synthesis, a ``sol'' (a colloidal solution of particles in a solvent) is first heated to form a ``gel'', which is then typically heated in multiple consecutive steps to form the desired product \citep{Vioux1997, Danks2016}. Common heating steps in both synthesis methods include calcination, where a mixture of compounds is heated to a high temperature to remove impurities and unwanted volatile substances, often through thermal decomposition \citep{west2014solid}, and sintering, where a compound is heated at a temperature (often higher than that reached by calcination) to induce nucleation and grain growth \citep{west2014solid}. 

% In annealing, a material is heat-treated above its recrystallization temperature to allow for increased diffusion of atoms, decreasing dislocation density, and increasing strain-free grain growth. Finally, drying involves heating at a relatively low temperature to remove solvent impurity through evaporation. In the domain of inorganic synthesis, knowledge of the appropriate experimental heating temperatures and times is crucial, as these conditions can control the identity, purity, and quality of the target phase(s) formed during the synthesis.

\section{Methods}
Our dataset consists of two publicly released materials synthesis databases (in JSON format) text-mined from scientific literature using a combination of NLP and rule-based extraction techniques \citep{Kononova2019}. The first is a solid-state synthesis database \citep{Kononova2019} containing 31,782 inorganic solid-state chemical reactions, while the second is a sol-gel synthesis database \citep{Kononova2019} containing 9,518 inorganic sol-gel chemical reactions. Each entry in the synthesis database contains a \textit{target material:} the stoichiometric formula of the target compound synthesized in the reaction, \textit{precursor materials:} the starting materials reacted together to form the target material, where a precursor is defined as a compound which shares one or more elements with the target material, excluding abundant elements that can be found in air (e.g. oxygen, hydrogen), and \textit{processing actions and synthssis conditions:} the sequence of synthesis actions (e.g. mix, grind, calcine, sinter, dry) that were performed on the precursor materials to transform them into the target material. Relevant synthsis conditions for these processing actions include temperatures and times (when reported in the synthesis). The datasets report an overall extraction accuracy of 93\% \citep{Kononova2019}.

\begin{figure}[h]
  \centering
  \includegraphics[width=10cm]{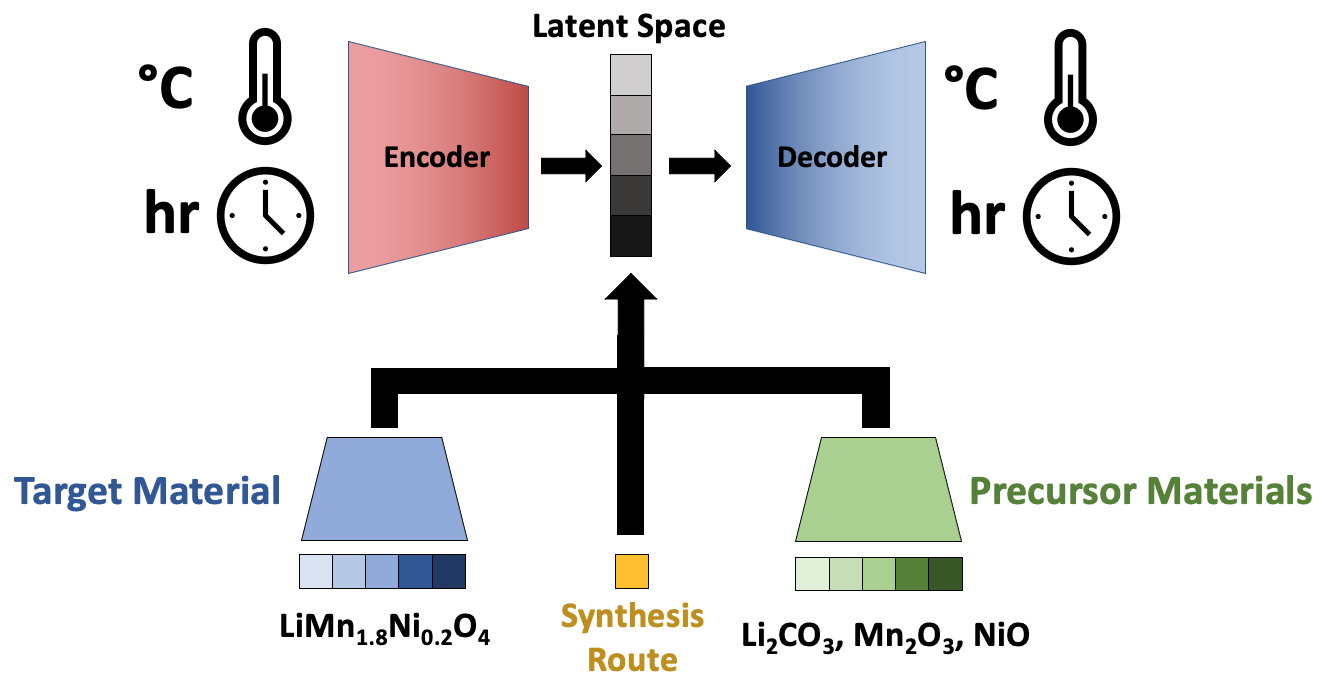}
  \caption{Architecture of CVAE used for temperature and time generation. The encoder embeds the synthesis conditions $x$ to a latent vector $z$, and the decoder reconstructs the synthesis conditions from the latent vector. The latent vector is concatenated with learned representations of the target and precursor materials as well as a binary indicator of the synthesis route.}
  \label{fig:fig5}
\end{figure}

 To make predictions for synthesis conditions we used (see Appendix) a conditional variational autoencoder (CVAE) with a convolutional encoder and recurrent decoder. A depiction of the model architecture is shown in Fig. \ref{fig:fig5}. Temperatures and times of the four heating steps of interest (calcination, sintering, annealing, and drying) were represented as an 8-dimensional vector and standardized. Targets and precursors (represented by their chemical formulas, see Appendix) were used as conditions and encoded and concatenated with the latent space representation using convolutional layers over sequences of one-hot vectors, where the total vocabulary is a character set consisting of the different elements and the numerical digits. Two dataset splits were investigated: a random split, and a compositional-based split based on \citep{Bartel2020} where the train, validation, and test set do not contain materials involving the same set of elements. For instance, if \ch{LiFePO4} is in the test set, then no other materials in the Li-Fe-P-O phase system (such as \ch{LiFeP2O7}) would be allowed in the training or validation sets.
 
\begin{figure}[h]
  \centering
  \includegraphics[width=14.5cm]{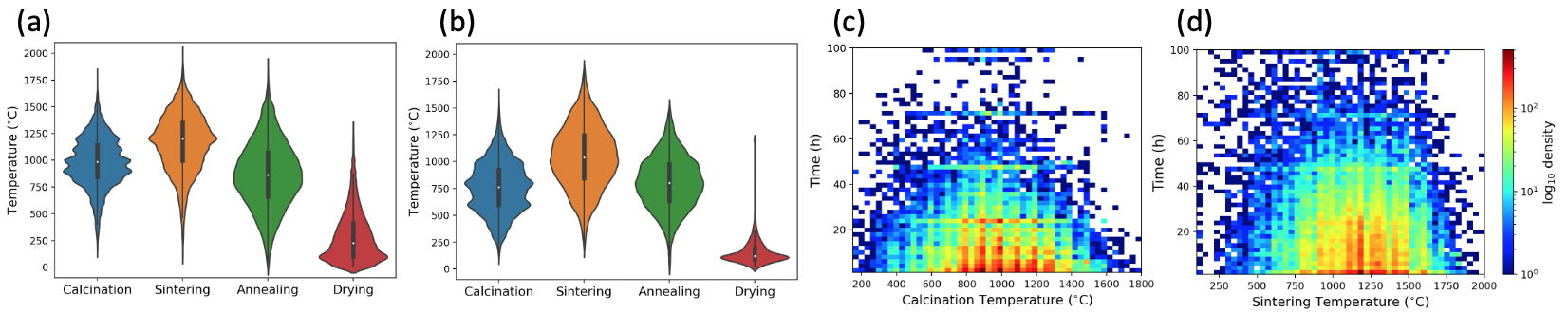}
  \caption{Temperature distributions for extracted operations for (a) solid-state and (b) sol-gel synthesis methods. (c) Calcination and (d) sintering time-temperature distributions for solid-state synthesis.}
  \label{fig:fig1}
\end{figure}

As shown in Fig. \ref{fig:fig1}, we plot the overall distribution of calcination, sintering, annealing, and drying temperatures and times for both solid-state and sol-gel synthesis methods. The dataset is mainly composed of unique compounds, with the majority of the entries reporting novel target compositions. Notably, calcination temperatures follow a common trend such that nitride $>$ oxide $>$ carbonate $>$ nitrate $>$ acetate, alkoxide, acetylacetonate, oxalate which is correlated to bonding strength between cations and anions \cite{He2020}. However, these trends are not absolute, as other factors such as decomposition reactions, reactivity of precursors, and identity of the target material play a role in the chosen calcination temperature.

\section{Results and Discussion}

\begin{figure}[h]
  \centering
  \includegraphics[width=14.5cm]{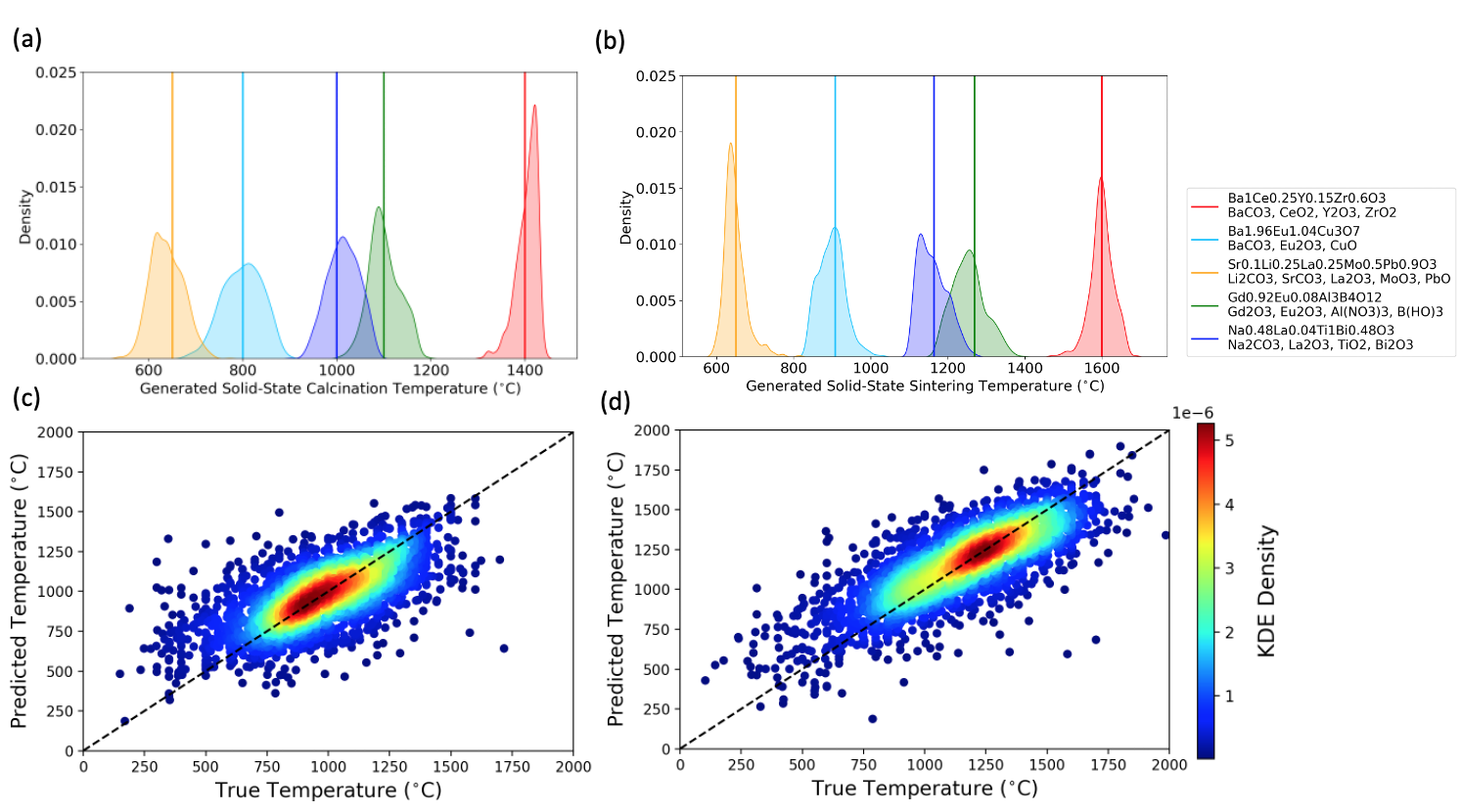}
  \caption{Generated solid-state (a) calcination and (b) sintering temperature distributions for a selection of samples in the held-out test set. Parity plots for predicted vs. true means of test set distributions of (c) calcination and (d) sintering temperatures. In the key, the target material is presented first (above) and precursor materials are presented second (below) for each entry.}
  \label{fig:fig2}
\end{figure}

In Fig. \ref{fig:fig2}, we plot the predicted calcination and sintering temperatures for a selection of samples in the held-out test set. The CVAE model indeed learns a meaningful relationship between the composition of the target material, the identities of the precursors in the synthesis, and a range within which the appropriate temperature for these heating operations should occur. For instance, in Fig. \ref{fig:fig2} (a)-(b) it is evident that when a compound consisting primarily of barium is doped with cesium, yttrium, and zirconium, it should require higher calcination and sintering temperatures than if doped with europium and copper. Moreover, the CVAE learns the common synthesis trend that sintering temperatures tend to be higher than calcination temperatures by the order of 100-300$^{\circ}$C. In Fig. \ref{fig:fig2} (c)-(d), we show the parity plots for predicted vs. true mean calcination and sintering temperatures in the test set.

The quantitative performance metrics of the CVAE model are presented in Table \ref{table1}, including mean absolute error (MAE), root-mean-square error (RMSE), mean relative error (MRE), and coefficient of determination ($R^{2}$). In experimental synthesis, suggesting approximate initial temperatures would be sufficiently helpful to inform experiments and accelerate synthesis of novel materials. On a held-out test set, the CVAE predicts the mean calcination temperature with a mean absolute error of 132.4 $^{\circ}$C and mean sintering temperature with a MAE of 129.9 $^{\circ}$C. To evaluate model performance against physically meaningful baselines, we leverage two common heuristics in the materials synthesis field as predictors on the dataset. Tamman's rule \citep{Merkle2005} is a synthesis heuristic which approximates reaction temperature as two-thirds the melting point of the lowest melting temperature reactant. For sintering temperature, we use a heuristic which approximates the value as 200 $^{\circ}$C above the calcination temperature predicted by Tamman's rule, which is the average difference between sintering and calcination temperatures in the dataset. From Table \ref{table1}, it is clear that the CVAE outperforms both heuristics by a factor of three to four.

\begin{table}[h]
  \caption{Dataset-wide performance of CVAE model compared to baseline heuristics}
  \label{table1}
  \centering
  \begin{tabular}{lllllr}
    \toprule
    Model               & Prediction Task   & MAE ($^{\circ}$C)  & RMSE ($^{\circ}$C)  & MRE (\%)   & $R^{2}$ \\
    \midrule
    CVAE (random split) & Mean Calcination Temp. & 132.4 & 180.9  & 16 & 0.40     \\
    CVAE (comp. split)  & Mean Calcination Temp  & 147.0  & 190.9  & 18  & 0.32   \\
    Tamman's Rule         & Mean Calcination Temp. & 679.2 & 596.8   & 60  & -8.68     \\
    \midrule
    CVAE (random split) & Mean Sintering Temp. & 129.9  & 173.4 & 13 & 0.60     \\
    CVAE (comp. split)  & Mean Sintering Temp  & 147.4   & 191.8 & 15  & 0.43    \\
    Sintering Heuristic & Mean Sintering Temp. & 529.0  & 613.9  & 45  & -5.03   \\
    \bottomrule
  \end{tabular}
\end{table}

\begin{figure}[h]
  \centering
  \includegraphics[width=14cm]{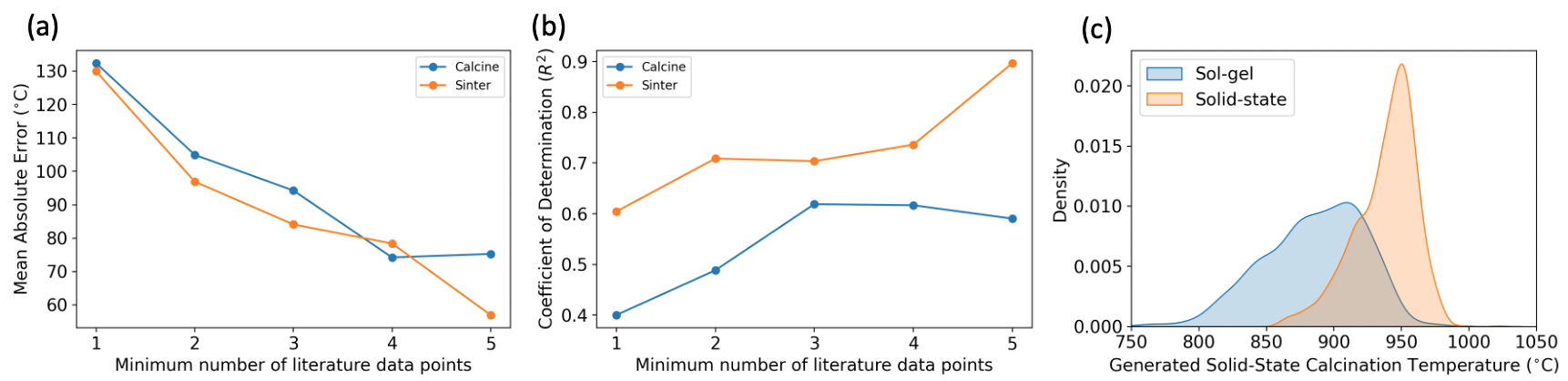}
  \caption{(a) MAE and (b) $R^{2}$ metrics as a function of minimum number of literature data points in the test set. (c) Generated calcination temperature distributions conditioned on either sol-gel or solid-state synthesis routes.}
  \label{fig:fig3}
\end{figure}

Another factor to consider in evaluating model performance is the structure of the dataset itself. In scientific literature, the majority of publications report the unique synthesis of a single compound or family of compounds, meaning the reported synthesis temperature may not be representative of either the optimal temperature or the range of temperature within which it is feasible to synthesize the compound. Thus, since the dataset is comprised of single experimental data points per material, comparing the means of generated distributions of temperatures to single experimental points likely underestimates model performance. For instance, in Fig. \ref{fig:fig3} (a)-(b), we plot the MAE and $R^{2}$ metrics as a function of the minimum number of literature data points for each example in the test set, showing marked improvement as the minimum number of points increases from one to five. With the experimental mean comprised of at least five points, the MAE is 57.0 $^{\circ}$C and $R^{2}$ is 0.90 for sintering and 75.3 $^{\circ}$C and 0.59 for calcination, respectively. We note that the MAE for calcination increases slightly as the minimum number of literature data points increases from 4 to 5 and the $R^{2}$ for calcination decreases slightly as the minimum number of literature data points increases from 3 to 5, which can be attributed to anthropomorphic factors in data reporting and systematic error discussed later.

\begin{figure}[h]
  \centering
  \includegraphics[width=14cm]{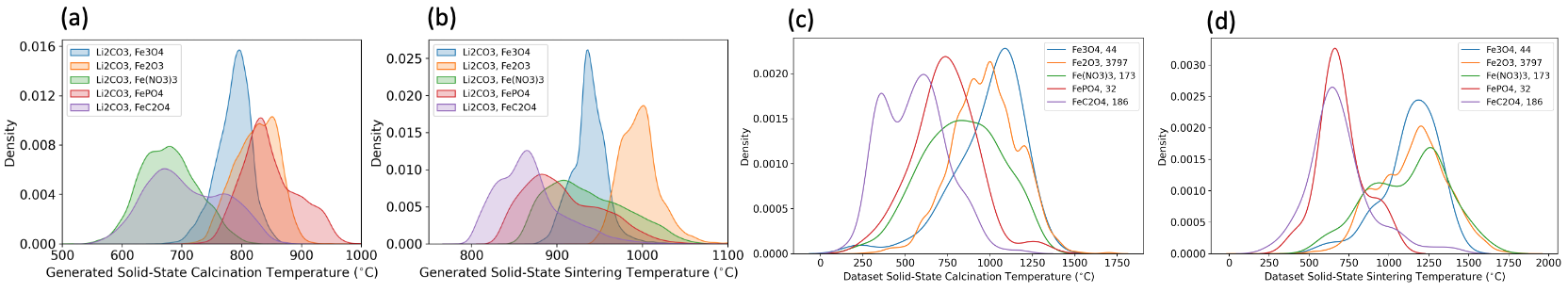}
  \caption{Generated solid-state (a) calcination and (b) sintering temperatures for the synthesis of \ch{Li4Fe7O12} from \ch{Li2CO3} and various iron-containing precursors. (c) Calcination and (d) sintering temperature trends as a function of iron-containing precursors across the dataset. The inset number is the frequency of the reported precursor in the dataset.}
  \label{fig:fig4}
\end{figure}

To show that the CVAE also learns the appropriate trends in synthesis route and precursor substitutions, we plot in Fig. \ref{fig:fig3}(c) the predicted calcination temperature distributions for the synthesis of \ch{Sr2FeO4} from \ch{Sr(NO3)2} and \ch{Fe(NO3)3}, conditioning on either sol-gel or solid-state synthesis. Evidently, the CVAE learns the relationship that if we change our choice of synthesis route from solid-state to sol-gel, the calcination and sintering temperatures should decrease as well, reflective of the calcination temperature distributions in both datasets. We also plot in \ref{fig:fig4}(a)-(b) the predicted calcination and sintering temperatures for the synthesis of \ch{Li4Fe7O12} from \ch{Li2CO3} while varying the identity of the iron-based precursor. The CVAE model recognizes that when we alter our choice of precursor from high bonding strength candidates such as \ch{Fe3O4} and \ch{Fe2O3} to lower bonding strength candidates such as \ch{Fe(NO3)3}, \ch{FePO4}, and \ch{FeC2O4}, the calcination and sintering temperatures should appropriately decrease. These trends are reflective of the overall shift in calcination and sintering temperature across the entire dataset portrayed in Fig. \ref{fig:fig4} (c)-(d). However, these trends are not absolute, as for example the generated calcination temperature distribution using \ch{FePO4} is higher than what would be expected simply by comparing with the literature-wide trends. While reported trends in synthesis temperatures and average bonding strength do correlate to an extent, other factors such as reactivity of precursors, intermediate reactions such as decomposition, the identity of the target compound, errors in automated extraction, and anthropomorphic factors \citep{He2020, Jia2019} such as experimentalist bias and past reported literature success all affect reported reaction conditions and influence the learned distributions.
 
\section{Conclusions}
We propose a CVAE model which suggests appropriate distributions for calcination and sintering temperatures in inorganic synthesis based on synthesis route, precursor identity, and the desired target compound. The model captures physics-based trends in the doping of target materials as well as choice of precursor and significantly outperforms common heuristics in the field in suggesting predictions of synthesis temperatures. We envision this model as a stepping-stone to high-throughput inorganic synthesis, where laboratory experiments will ultimately be informed by machine learning to hasten the synthesis and optimization of inorganic materials with desirable properties.

\section{Broader Impact}
Inorganic materials play a crucial role in the advancement of fields such as energy storage, chemical catalysis, thermoelectrics, and microelectronics. Therefore, new computational tools leveraging machine learning and scientific data need to be developed in order to accelerate the design and realization of novel materials with desirable properties. Our approach guides experimentalists with initial suggestions to aid in synthesis planning for the discovery and design of new materials. Outside of inorganic materials science, our approach can potentially be applied to a wide range of other fields which are currently adapting machine learning-based tools to unify theoretical predictions and experimental validation, such as drug discovery and electronic device manufacturing.

\section{Acknowledgements}
This material is based upon work supported by the National Science Foundation Graduate Research Fellowship under Grant No. 1745302. The information, data, or work presented herein was also funded in part by the Advanced Research Projects Agency-Energy (ARPA-E), U.S. Department of Energy, under Award Number DE-AR0001209. The views and opinions of authors expressed herein do not necessarily state or reflect those of the United States Government or any agency thereof.

\bibliography{references}

\section{Appendix}

\subsection{Theory}
In a variational autoencoder (VAE), the loss function consists of the variational lower bound, also known as the evidence lower bound (ELBO):

\begin{equation}
    \mathcal{L}_{\theta, \phi} = \mathbb{E}_{q_{\phi}(z|x)}[\log p_{\theta}(x|z)] - D_{KL}(q_{\phi}(z|x) \Vert p_{\theta}(z))
\end{equation}

where $\mathbb{E}$ is the expectation value, $p$ and $q$ are probability distributions, $D_{KL}$ is the Kullback–Leibler divergence, and $x$ and $z$ are the data and latent spaces, respectively. The first and second terms are often called the reconstruction loss and the KL loss, respectively. The reconstruction loss encourages the decoder to learn to reconstruct the data, while the KL loss is a regularization term which measures how similar the variational distribution encoded from the input data $q_{\phi}(z|x)$ and the latent space distribution $p_{\theta}(z)$ are. We take $p_{\theta}(z)$ to be a standard normal distribution with zero mean and unit variance, such that $p_{\theta}(z) = \mathcal{N}(0, 1)$. In a VAE, $q_{\phi}(z|x)$ and $p_{\theta}(x|z)$ are approximated by an encoder and a decoder, respectively. For a conditional variational autoencoder (CVAE), we embed the conditional information as a vector (denoted by $c$) in the objective function of the VAE, leading to the revised loss function as follows:

\begin{equation}
    \mathcal{L}_{\theta, \phi} = \mathbb{E}_{q_{\phi}(z|x)}[\log p_{\theta}(x|z,c)] - D_{KL}(q_{\phi}(z|x,c) \Vert p_{\theta}(z|c))
\end{equation}

In our model architecture, the data $x$ consists of standardized temperatures and times of the four heating steps of interest (calcination, sintering, annealing, and drying) represented as an 8-dimensional vector. The condition $c$ consists of targets and precursors (represented by their chemical formulas), encoded and concatenated with the latent space representation using convolutional layers over sequences of one-hot vectors, where the total vocabulary is a character set consisting of the different elements and the numerical digits. An additional condition (if desired) is the synthesis route, which is a binary condition (solid-state or sol-gel). Using this framework, the CVAE model can generate temperature and time distributions conditioned on the target and precursor materials of interest.

\subsection{Model architecture}
 All neural network models were implemented in the Keras library using the TensorFlow backend. For the conditional variational autoencoder (CVAE) model, convolutional layers encode the temperature-time vector into a latent parameter space for means and variances of Gaussian variational posteriors, and outputs from a latent sampling function are concatenated with conditional inputs as inputs to a recurrent decoder. The encoder is comprised of three convolutional layers and the decoder comprised of three gated recurrent unit (GRU) layers. In producing the results for this study, 3 latent dimensions were used. Targets and precursors (represented by their chemical formulas) were used as conditions and encoded and concatenated with the latent space representation using convolutional layers over sequences of one-hot vectors, where the total vocabulary is a character set consisting of the different elements and the numerical digits. A period was included in the character set for targets to represent non-stoichiometric target formulas. Training was conducted on two NVIDIA Titan Xp GPUs with a batch size of 128 and the Adam optimizer with default hyperparameters. Hyperparameter selection was performed by grid searches, where the latent layer dimension was varied from 2 to 5 dimensions and the standard deviation of the Gaussian prior was varied between 0.001 and 10.0. The data was split in a 75/15/10 train/validation/test ratio using either random or compositional splits, and hyperparameters were selected based on minimizing validation loss.

\subsection{Data post-processing}
Our post-processing of the data was completed as follows. First, reactions with organic precursors and targets, non-stoichiometric precursors, unsubstituted target stoichiometries, less than two or greater than five precursors, or not containing at least one relevant heating step with a reported temperature were removed. Hydrate precursors were truncated to their base chemical formula. Temperatures and times were converted into units of Celsius and Hours and limited to between 100 $^{\circ}$C and 2000 $^{\circ}$C and less than 100 hours, and if an operation step was reported with more than one temperature or time, the highest value was taken. If a relevant heating step occurred more than once in a recipe, the last value was taken. Because not every synthesis recipe employed all four heating steps, data imputation was conducted using the \texttt{IterativeImputer} module in \texttt{scikit-learn} with the BayesianRidge estimator, default hyperparameters, and minimum and maximum imputation values set to those in the dataset. To aid in accurate imputation, precursors were one-hot encoded and used as additional features. Temperatures and times were standardized by removing the mean and scaling to unit variance per feature.

%and target materials were featurized using Magpie \citep{Ward2016} embeddings from the Matminer package.

%%%%%%%%%%%%%%%%%%%%%%%%%%%%%%%%%%%%%%%%%%%%%%%%%%%%%%%%%%%%
\section*{Checklist}

% %%% BEGIN INSTRUCTIONS %%%
% The checklist follows the references.  Please
% read the checklist guidelines carefully for information on how to answer these
% questions.  For each question, change the default \answerTODO{} to \answerYes{},
% \answerNo{}, or \answerNA{}.  You are strongly encouraged to include a {\bf
% justification to your answer}, either by referencing the appropriate section of
% your paper or providing a brief inline description.  For example:
% \begin{itemize}
%   \item Did you include the license to the code and datasets? \answerYes{}%See Section~\ref{gen_inst}.}
%   \item Did you include the license to the code and datasets? \answerNo{The code and the data are proprietary.}
%   \item Did you include the license to the code and datasets? \answerNA{}
% \end{itemize}
% Please do not modify the questions and only use the provided macros for your
% answers.  Note that the Checklist section does not count towards the page
% limit.  In your paper, please delete this instructions block and only keep the
% Checklist section heading above along with the questions/answers below.
% %%% END INSTRUCTIONS %%%

\begin{enumerate}

\item For all authors...
\begin{enumerate}
  \item Do the main claims made in the abstract and introduction accurately reflect the paper's contributions and scope?
    \answerYes{}
  \item Did you describe the limitations of your work?
    \answerYes{}
  \item Did you discuss any potential negative societal impacts of your work?
    \answerNA{} Our work focuses on leveraging scientific data from literature to advance materials discovery, so there is no potential negative societal impact to discuss.
  \item Have you read the ethics review guidelines and ensured that your paper conforms to them?
    \answerYes{}
\end{enumerate}

\item If you are including theoretical results...
\begin{enumerate}
  \item Did you state the full set of assumptions of all theoretical results?
    \answerNA{}
	\item Did you include complete proofs of all theoretical results?
    \answerNA{}
\end{enumerate}

\item If you ran experiments...
\begin{enumerate}
  \item Did you include the code, data, and instructions needed to reproduce the main experimental results (either in the supplemental material or as a URL)?
    \answerYes{The data is open-source and freely available. The code and project are still a work-in-progress and code will be released upon full publication of this work at a later date.}
  \item Did you specify all the training details (e.g., data splits, hyperparameters, how they were chosen)?
    \answerYes{}
	\item Did you report error bars (e.g., with respect to the random seed after running experiments multiple times)?
    \answerNA{}
	\item Did you include the total amount of compute and the type of resources used (e.g., type of GPUs, internal cluster, or cloud provider)?
    \answerYes{}
\end{enumerate}

\item If you are using existing assets (e.g., code, data, models) or curating/releasing new assets...
\begin{enumerate}
  \item If your work uses existing assets, did you cite the creators?
    \answerYes{}
  \item Did you mention the license of the assets?
    \answerYes{}
  \item Did you include any new assets either in the supplemental material or as a URL?
    \answerNA{}
  \item Did you discuss whether and how consent was obtained from people whose data you're using/curating?
    \answerNA{}
  \item Did you discuss whether the data you are using/curating contains personally identifiable information or offensive content?
    \answerNA{}
\end{enumerate}

\item If you used crowdsourcing or conducted research with human subjects...
\begin{enumerate}
  \item Did you include the full text of instructions given to participants and screenshots, if applicable?
    \answerNA{}
  \item Did you describe any potential participant risks, with links to Institutional Review Board (IRB) approvals, if applicable?
    \answerNA{}
  \item Did you include the estimated hourly wage paid to participants and the total amount spent on participant compensation?
    \answerNA{}
\end{enumerate}

\end{enumerate}

\end{document}